\documentclass[12pt]{article}

\usepackage{color}
\usepackage{a4}
\usepackage{latexsym}
\usepackage{cite}
\usepackage{lineno}

\usepackage{array}
\newcolumntype{T}{>{\ttfamily} c}
\newcolumntype{M}{>{$\displaystyle} c <{$}}

\usepackage{epsfig}
\usepackage{graphicx}
\usepackage{hyperref}

\usepackage{pslatex}
\usepackage[latin1]{inputenc}
\usepackage[T1]{fontenc}

\usepackage[dvipsnames]{xcolor}
\usepackage{color,colordvi}
\def\colour4colour#1{\Blue{#1}}

\def\slash#1{\rlap{\hbox{$\mskip 1 mu /$}}#1}      % good slash for lower case
\newcommand{\beq}{\begin{equation}}
\newcommand{\eeq}{\end{equation}}
\newcommand{\bea}{\begin{eqnarray}}
\newcommand{\eea}{\end{eqnarray}}
\newcommand{\nn}{\nonumber}
\newcommand{\MSb}{$\overline{\mbox{MS}}$}

\newcommand{\als}{\alpha_{\rm s}}
\newcommand{\ars}{a_{\rm s}}
\newcommand{\ep}{\varepsilon}

\newcommand{\hspn}{{\hspace{-2mm}}}
\newcommand{\hspp}{{\hspace{3mm}}}

\def\frct#1#2{\mbox{\small{$\displaystyle\frac{#1}{#2}$}}}

\def\as(#1){{\alpha_{\rm s}^{\,#1}}}
\def\ar(#1){{a_{\rm s}^{\,#1}}}

\def\mus{{\mu^{\,2}}}

\def\B(#1,#2){{\beta_{#1}^{\,#2}}}

\def\nc{{n_c}}
\def\ca{{C^{}_A}}

\def\cf{{C^{}_F}}
\def\cfs{{C^{\, 2}_F}}

\def\nf{{n^{}_{\! f}}}

\def\nfs{{n^{\,2}_{\! f}}}
\def\nft{{n^{\,3}_{\! f}}}

\def\pgq{ {p_{gq}} }
\def\pxgq{ {p_{gq}(x)} }
\def\pmxgq{ {p_{gq}(-x)} }
\def\ptmx{ {(2-x)} }

\def\triv#1{#1}

\def\gqg0N{p_{\rm qg}^{}}

\def\xm1{{(1 \! - \! x)}}
\def\xp1{{(1 \! + \! x)}}

\def\Lnt(#1){\ln^{\,#1}(1\!-\!x)}
\def\pqq(#1){p_{\rm{qq}}(#1)}

\def\z#1{{\zeta_{#1}}}

\def\col{\color{RubineRed}}

\def\Ndelta{{\col\delta(N\!-\!2)}}
\def\th{{\col\theta(N\!-\!4)}}
\def\St(#1){{\col \mathbf S_{#1}(N\!-\!2)}}

\def\S(#1){{\col \mathbf S_{#1}}}
\def\Ss(#1,#2){{\col{ \mathbf S}_{#1,#2}}}
\def\Sss(#1,#2,#3){{\col{ \mathbf S}_{#1,#2,#3}}}
\def\Ssss(#1,#2,#3,#4){{\col{ \mathbf S}_{#1,#2,#3,#4}}}
\def\Sssss(#1,#2,#3,#4,#5){{\col{ \mathbf S}_{#1,#2,#3,#4,#5}}}
\def\Ssssss(#1,#2,#3,#4,#5,#6){{\col{ \mathbf S}_{#1,#2,#3,#4,#5,#6}}}

\def\H(#1){{\col{\rm{H}}_{#1}}}
\def\Hh(#1,#2){{\col{\rm{H}}_{#1,#2}}}
\def\Hhh(#1,#2,#3){{\col{\rm{H}}_{#1,#2,#3}}}
\def\Hhhh(#1,#2,#3,#4){{\col{\rm{H}}_{#1,#2,#3,#4}}}
\def\Hhhhh(#1,#2,#3,#4,#5){{\col{\rm{H}}_{#1,#2,#3,#4,#5}}}
\def\Hhhhhh(#1,#2,#3,#4,#5,#6){{\col{\rm{H}}_{#1,#2,#3,#4,#5,#6}}}

\def\Sp(#1,#2){{{S}_{#1}^{\,#2}}}

\def\Dplus(#1){\mathcal{D}_{#1}}
\def\D(#1){{ D_{#1}}}
\def\Dd(#1,#2){{ D_{#1}^{\,#2}}}
\def\Lmzp^#1{\ln^{\,#1}(1\!-\!x)}
\def\Lmz{\ln(1\!-\!x)}
\def\Lz{\ln x}
\def\Lzp^#1{\ln^{\,#1\!} x} 

\def\frct#1#2{\mbox{\large{$\frac{#1}{#2}$}}}

\begin{document}
\setlength{\parskip}{0.2cm}
\setlength{\baselineskip}{0.55cm}

% --------------------------------------------------------------------

\begin{titlepage}
\noindent
DESY 23--146 \hfill October 2023\\
LTH 1353 \\
\vspace{0.6cm}
\begin{center}
{\LARGE \bf The double fermionic contribution to the \\[1ex]
 four-loop quark-to-gluon splitting function}\\ 
\vspace{2.0cm}
\large
G.~Falcioni$^{\, a,b}$, F.~Herzog$^{\, c}$, S. Moch$^{\, d}$,
J. Vermaseren$^{\, e}$ and A. Vogt$^{\, f}$\\

\vspace{1.2cm}
\normalsize
{\it $^a$Dipartimento di Fisica, Universit\`{a} di Torino,
  Via Pietro Giuria 1, 10125 Torino, Italy}\\
\vspace{1mm}
{\it $^b$ Physik-Institut, Universit\"{a}t Z\"{u}rich, 
  Winterthurerstrasse 190, 8057 Z\"{u}rich, Switzerland}\\
\vspace{5mm}
{\it $^c$Higgs Centre for Theoretical Physics, School of Physics and Astronomy\\
  The University of Edinburgh, Edinburgh EH9 3FD, Scotland, UK}\\
\vspace{5mm}
{\it $^d$II.~Institute for Theoretical Physics, Hamburg University\\
\vspace{0.5mm}
Luruper Chaussee 149, D-22761 Hamburg, Germany}\\
\vspace{4mm}
{\it $^e$Nikhef Theory Group, 
  Science Park 105, 1098 XG Amsterdam, The Netherlands} \\
\vspace{4mm}
{\it $^f$Department of Mathematical Sciences, University of Liverpool\\
\vspace{0.5mm}
Liverpool L69 3BX, United Kingdom}\\
\vspace{2.4cm}
{\large \bf Abstract}
\vspace{-0.2cm}
\end{center}
We have computed the first 30 even-$N$ moments for the double fermionic
($\nfs\,$) part of the quark-to-gluon splitting function $P_{\rm gq}$ 
at the fourth order of perturbative QCD via the renormalization of 
off-shell operator matrix elements.
From these results we have determined the all-$N$ form, and hence
the exact $x$-space expression, using systems of Diophantine equations 
for its coefficients. 
The dominant and subdominant leading small-$x$ $\nfs$ contributions to
$P^{\,(3)}_{\rm gq}(x)$ are of the form $x^{\,-1} \ln x$ and 
$\ln^{\,4\!} x$, respectively; the leading large-$x$ term is 
$\ln^{\,4\!} \xm1$. 
The coefficient of the first of these is new, the other two agree with 
results obtained before and thus provide checks of our results. 

\vspace*{0.5cm}
\end{titlepage}

\newpage
% ---------------------------------------------------------------------

Next-to-next-to-next-to-leading order (N$^3$LO) calculations of 
benchmark  processes in perturbative QCD form an important part of 
the accuracy frontier at the Large Hadron Collider LHC.
Complete analyses at this order require the 4-loop splitting functions
for the scale dependence of the proton's parton distributions 
in the fractional momentum $x$,
in the flavour singlet sector given~by
\beq
\label{eq:sgEvol}
 \frac{d}{d \ln\mus} \;
 \Big( \begin{array}{c}
         \! q_{\rm s}^{} \!\! \\ \!g\! \end{array} \Big)
 \: = \: \left( \begin{array}{cc}
         \! P_{\rm qq} & P_{\rm qg} \!\!\! \\
         \! P_{\rm gq} & P_{\rm gg} \!\!\! \end{array} \right)
 \otimes
 \Big( \begin{array}{c}
         \! q_{\rm s}^{}\!\! \\ \!g\!  \end{array} \Big)
 \quad \mbox{with} \quad
 P_{\,\rm ik}^{}(x,\als)
    \,=\, \sum_{n=0} \ar(n+1)\,P_{\,\rm ik}^{\,(n)}(x)
 \: ,
\eeq
where
$q_{\rm s}^{} \,=\, \sum_{\,i=1}^{\,\nf} \, ( q_i^{} + \bar{q}_i^{} )$
and $g$ are the singlet quark and gluon distributions and $\otimes$
denotes the Mellin convolution in the momentum variable $x$.
The determination of these splitting functions requires very involved 
calculations, and only partial results have been obtained so~far.

The leading $\nft$ contributions in the limit of a large numbers of 
light flavours $\nf$ were completed in ref.~\cite{Davies:2016jie}
together with the $\nfs$ parts of the flavour non-singlet quark-quark 
splitting functions $P_{\,\rm ns}^{\,(3)\pm}(x)$. Very recently these
results have been extended to the $\nfs$ part of the pure-singlet
quark-quark splitting function $P_{\,\rm ps}^{\,(3)}(x) 
\,=\, P_{\,\rm qq}^{\,(3)}(x) - P_{\,\rm ns}^{\,(3)+}(x)$
\cite{Gehrmann:2023cqm}. The N$^3$LO non-singlet splitting functions 
are completely known in the limit of a large number of colours $\nc$
\cite{Moch:2017uml}.

Beyond the quark-quark cases and the $\nft$ terms, only a
limited number of moments of the 4-loop splitting functions have 
been computed so far. 
These quantities correspond (up to a conventional sign) to the 
anomalous dimensions of the gauge invariant operators of twist two,
\beq
\label{eq:ggq3N}
\gamma_{\,\rm ik}^{\,(3)}(N) \:=\: - \int_0^1 \!dx\:\, x^{\,N-1}\,
P_{\,\rm ik}^{\,(3)}(x).
\eeq
Building on recent progress on the renormalization of flavour-singlet 
operator matrix elements (OMEs), in particular 
refs.~\cite{Falcioni:2022fdm,Gehrmann:2023ksf}, we have been able
to obtain the even-$N$ Mellin moments to $N=20$ of $P_{\rm ps}^{\,(3)}$
and $P_{\rm qg}^{\,(3)}$ \cite{Falcioni:2023luc,Falcioni:2023vqq}.
For the lower-row quantities in eq.~(\ref{eq:sgEvol}),
only determinations to $N\!=\!10$ have been completed so far, using the 
conceptually much easier but computationally harder route via structure
functions in deep-inelastic scattering (DIS) 
\cite{Moch:2021qrk,MRUVV-tba}.
In the present letter, we present the first analytic $x$-dependence 
of a 4-loop $\nfs$ contribution to eq.~(\ref{eq:sgEvol}) beyond the 
quark-quark case. 

As in refs.~\cite{Falcioni:2023luc,Falcioni:2023vqq}, our calculations
are performed in the framework of the operator-product expansion, via
off-shell OME's.
Therefore the renormalization of the results requires counterterms 
associated with gauge-variant and ghost operators, commonly known as
aliens.
However, only a few alien operators enter for the $\nfs$ contribution
in the quark-to-gluon case. 
In particular, this contribution always includes a factor $C_F$, hence 
only the colour factors $\cfs\, \nfs$ and $C_F C_A\,\nfs$ can appear 
at four loops. 
The former contribution, which is also present in QED, does not require 
any alien counterterms. 
The latter colour factor requires only one class of aliens, namely
(see ref.~\cite{Falcioni:2023luc})
\beq
\label{eq:alienI}
 O_{\rm q}^{\,I} \,=\, 
    \eta\, g\, \overline{\psi}\,\slash{\Delta}\, 
    t^a\, \psi\, \left(\partial^{\,N-2}A_a \right)
\, , \quad
 O_{\rm g}^{\,I} \,=\,
    \eta\, \left(D.F\right)^a \left(\partial^{\,N-2}A_a\right) 
\, , \quad
 O_{\rm c}^{\,I} \,=\,
    - \eta\, \left(\partial \,\bar{c}^{\,a}\right) 
      \left(\partial^{\,N-1}c_a \right)
\, .
\eeq
where $\psi$, $A$ and $c$ represent the quark, gluon and ghost fields, 
$F$ the gluon field-strength tensor and $D = \partial-{\rm i}gA$
the covariant derivative with the coupling $g$, where
$g^2/(4\pi) = \als = 4\pi\, \ars $, after contraction of the Lorentz
indices with $N$ identical light-like vectors $\Delta^{\,\mu}$.
The mixing $\eta$, which is a function of $N$ and $\als$, needs to be 
determined to three loops. This is achieved efficiently by 
renormalizing the 3-loop ghost-antighost correlator with an insertion 
of the gauge-invariant gluon operators. 
The results to $N=60$ agree with all-$N$ the expressions of 
ref.~\cite{Gehrmann:2023ksf}.

The Feynman diagrams contributing to the $\nfs$ part of 
$\gamma_{\,\rm gq}^{\,(3)}(N)$
form a small subset of those for the general case. A few examples 
are shown in fig.~1. The diagrams are processed in the usual manner, 
see ref.~\cite{Herzog:2016qas}, by a {\sc Form} 
\cite{Vermaseren:2000nd,Kuipers:2012rf,Ruijl:2017dtg} program
which collects self-energy insertions, determines the colour factors 
\cite{vanRitbergen:1998pn} and classifies the topologies according 
to the conventions of the {\sc Forcer} program~\cite{Ruijl:2017cxj}.
An optimized in-house version of this program has been employed
to perform the integral reduction for fixed even values of $N$ in
$4-2\ep$ dimensions.
For the highest moments of the hardest 3-loop diagrams we have used
the {\sc Mincer} program \cite{Gorishnii:1989gt,Larin:1991fz}, 
suitably extended in its tables from the version optimized for 
ref.~\cite{Moch:2014sna}. 

\begin{figure}[t]
\centerline{\epsfig{file=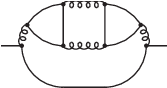,width=5.5cm,angle=0}%
\hspace{1cm}\epsfig{file=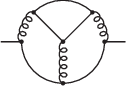,width=4cm,angle=0}}
\caption{\label{fig1} \small
Sample diagrams that enter, after inserting the gluon operator into
one of the gluon lines, the computation of the $\nfs$ part of
$\gamma_{\,\rm gq}^{\,(3)}(N)$ at even values of $N$.
Left: the only genuine 4-loop diagram. Due to Furry's theorem its
colour factor is $C_F C_A \,\nfs$.
Right: the 3-loop and (after replacing one of the gluon lines by a
one-loop gluon propagator) 4-loop cases that lead to the hardest
computations at large $N$. }
\vspace*{-2mm}
\end{figure}

Using these tools, we have been able to compute the anomalous 
dimensions $\gamma_{\,\rm gq}^{\,(3)}(N)$ in eq.~(\ref{eq:ggq3N}) 
at all even values $ 2 \leq N \leq 60 $. 
The corresponding all-$N$ expressions include Riemann-$\zeta$ values, 
harmonic sums \cite{Vermaseren:1998uu} and simple denominators 
$\,D_a^{\,k} \equiv (N+a)^{-k}$. 
The range of $a$, the maximal powers of $D_a$ and maximal weights of 
the sums can be inferred from the prime-factor decomposition of the 
denominators at $N \leq 60$. 
Also lower-order expressions, in particular the $\cfs\,\nf$ and
$C_F C_A \,\nf$ parts of $\gamma_{\,\rm gq}^{\,(2)}(N)$ 
\cite{Vogt:2004mw}, provide useful information about the expected
analytic form of $\gamma_{\,\rm gq}^{\,(3)}(N)$. 

In this manner, we are led to an ansatz of 143 functions
for the non-$\zeta$ $\cfs\,\nfs$ contribution, hence there are far 
too many unknown coefficients for a direct determination from the 
calculated moments.  However, these coefficients are integer modulo 
some powers of 2 and 3.
Therefore the resulting system of equations can be turned into a
Diophantine system which requires far fewer equations than
unknowns and which can be addressed by number-theoretical techniques,
as, e.g., in refs.~\cite{axbAlg,Calc}, based on the LLL algorithm
\cite{Lenstra1982}.
This approach has been successfully applied to 3-loop and 4-loop
splitting functions before, e.g., in refs.~%
\cite{Velizhanin:2012nm,Moch:2014sna,Davies:2016jie,Moch:2017uml}. 

We have eliminated 19 coefficients -- of all functions with overall
weights $w\!=\!1$ and $w\!=\!2$, and some of $w\!=\!3$ -- which are 
mostly expected to be very large, and determined the remaining 124 
unknowns from the 11 equations for $2 \leq N \leq 60$, using the 
program of ref.~\cite{Calc}. 
17 of these 124 coefficients turned out to vanish. 
An analogous procedure was used for the $C_F C_A\,\nfs$ part.
We will comment on the differences between the two cases below.

The resulting expression for the $\nfs$ part of the N$^3$LO 
quark-to-gluon splitting function at all even $N \geq 2$ reads
\bea
\lefteqn{ \gamma_{\rm gq}^{\,(3)}(N)\Big|_{\nfs} \;=\; }
  \nn \\[-2mm] & & \mbox{\hspn}
%%START
%%
%%L texggq3cf2nf2 = 
  \frct{16}{27}\* \cfs\* \nfs \* \bigg( 
  \pgq(N) \* (
 -8\,\*\S(4)
 -16\,\*\Ss(1,3)
 -8\,\*\Ss(2,2)
 -8\,\*\Ss(3,1)
 +36\,\*\Sss(1,1,2)
 +28\,\*\Sss(1,2,1)
 +16\,\*\Sss(2,1,1)
%%STOP
\nn \\[-0.5mm] & & \mbox{\hspp}
%%START
 +16\,\*\Ssss(1,1,1,1) )
 +24\,\* (N^2 + N + 2) 
  \,\*\triv(\D(0)^2\,\*\D(1)^2\,\*\D(2)\,\*\Dd(-1,2))
  \,\* ( 8\,\* \S(-3) - 12\,\* \Ss(-2,1) - 12\,\* \Ss(1,-2) )
%%STOP
\nn \\[0.5mm] & & \mbox{\hspp}
%%START
 +\S(3)\,\*(276\,\*\D(-1)
 -860/3\,\*\D(0)
 +259/3\,\*\D(1)
 -64/3\,\*\D(2)
 -64\,\*\Dd(-1,2)
 -100\,\*\Dd(0,2)
 -48\,\*\Dd(0,3)
%%STOP
\nn \\[0.5mm] & & \mbox{\hspp}
%%START
 -26\,\*\Dd(1,2)
 +24\,\*\Dd(1,3))
 +\Ss(1,2)\,\*(-224\,\*\D(-1)
 +758/3\,\*\D(0)
 -361/3\,\*\D(1)
 +16/3\,\*\D(2)
 +16\,\*\Dd(-1,2)
%%STOP
\nn \\[0.5mm] & & \mbox{\hspp}
%%START
 +20\,\*\Dd(0,2)
 +24\,\*\Dd(0,3)
 +8\,\*\Dd(1,2)
 -12\,\*\Dd(1,3))
 +\Ss(2,1)\,\*(-500/3\,\*\D(-1)
 +586/3\,\*\D(0)
 -275/3\,\*\D(1)
%%STOP
\nn \\[0.5mm] & & \mbox{\hspp}
%%START
 +16/3\,\*\D(2)
 +16\,\*\Dd(-1,2)
 +20\,\*\Dd(0,2)
 +24\,\*\Dd(0,3)
 -12\,\*\Dd(1,3))
 +\Sss(1,1,1)\,\*(-148\,\*\D(-1)
 +364/3\,\*\D(0)
%%STOP
\nn \\[0.5mm] & & \mbox{\hspp}
%%START
 -409/6\,\*\D(1)
 -16/3\,\*\D(2)
 -16\,\*\Dd(-1,2)
 -22\,\*\Dd(0,2)
 -24\,\*\Dd(0,3)
 +45\,\*\Dd(1,2)
 +12\,\*\Dd(1,3))
%%STOP
\nn \\[0.5mm] & & \mbox{\hspp}
%%START
 +\S(-2)\,\*(-5504/9\,\*\D(-1)
 +504\,\*\D(0)
 +24\,\*\D(1)
 +752/9\,\*\D(2)
 +712/3\,\*\Dd(-1,2)
 +480\,\*\Dd(0,2)
%%STOP
\nn \\[0.5mm] & & \mbox{\hspp}
%%START
 +96\,\*\Dd(0,3)
 +96\,\*\Dd(1,2)
 -144\,\*\Dd(1,3)
 -32/3\,\*\Dd(2,2))
 +\S(2)\,\*(5432/9\,\*\D(-1)
 -1402/3\,\*\D(0)
 +73/6\,\*\D(1)
%%STOP
\nn \\[0.5mm] & & \mbox{\hspp}
%%START
 -944/9\,\*\D(2)
 -416/3\,\*\Dd(-1,2)
 -1600/3\,\*\Dd(0,2)
 -80\,\*\Dd(0,3)
 -120\,\*\Dd(0,4)
 -212/3\,\*\Dd(1,2)
 +130\,\*\Dd(1,3)
%%STOP
\nn \\[0.5mm] & & \mbox{\hspp}
%%START
 +48\,\*\Dd(1,4)
 -32\,\*\Dd(2,2))
 +\Ss(1,1)\,\*(121/9\,\*\D(-1)
 +2237/18\,\*\D(0)
 -2219/36\,\*\D(1)
 +424/9\,\*\D(2)
%%STOP
\nn \\[0.5mm] & & \mbox{\hspp}
%%START
 +52\,\*\Dd(-1,2)
 +257/3\,\*\Dd(0,2)
 +102\,\*\Dd(0,3)
 -299/2\,\*\Dd(1,2)
 -20\,\*\Dd(1,3)
 +12\,\*\Dd(1,4)
 +16/3\,\*\Dd(2,2))
%%STOP
\nn \\[0.5mm] & & \mbox{\hspp}
%%START
 +\S(1)\,\*(82801/54\,\*\D(-1)
 -65503/36\,\*\D(0)
 +31639/72\,\*\D(1)
 -3086/27\,\*\D(2)
 -2660/9\,\*\Dd(-1,2)
%%STOP
\nn \\[0.5mm] & & \mbox{\hspp}
%%START
 -4915/9\,\*\Dd(0,2)
 -1351\,\*\Dd(0,3)
 +14\,\*\Dd(0,4)
 -312\,\*\Dd(0,5)
 -2219/12\,\*\Dd(1,2)
 -2491/6\,\*\Dd(1,3)
 +85\,\*\Dd(1,4)
%%STOP
\nn \\[0.5mm] & & \mbox{\hspp}
%%START
 +156\,\*\Dd(1,5)
 -352/3\,\*\Dd(2,2)
 -224/3\,\*\Dd(2,3))
 -1094983/324\,\*\D(-1)
 +282613/72\,\*\D(0)
%%STOP
\nn \\[0.5mm] & & \mbox{\hspp}
%%START
 -125953/144\,\*\D(1)
 +16636/81\,\*\D(2)
 +19192/27\,\*\Dd(-1,2)
 +1540\,\*\Dd(0,2)
 +2766\,\*\Dd(0,3)
%%STOP
\nn \\[0.5mm] & & \mbox{\hspp}
%%START
 +515/3\,\*\Dd(0,4)
 +766\,\*\Dd(0,5)
 +2265/8\,\*\Dd(1,2)
 +6425/12\,\*\Dd(1,3)
 -4019/6\,\*\Dd(1,4)
 -256\,\*\Dd(1,5)
%%STOP
\nn \\[0.5mm] & & \mbox{\hspp}
%%START
 +228\,\*\Dd(1,6)
 +312\,\*\Dd(2,2)
 +1312/9\,\*\Dd(2,3)
 -64\,\*\Dd(2,4)
%%STOP
%*csm 
% correctected here   \,+\, \z3\,\* (
%%START
 \,+\, 4\*\z3\,\* (
 -6\,\*\pgq(N)\,\*\S(1)
 +52\,\*\D(-1)
%%STOP
\nn \\[0.5mm] & & \mbox{\hspp}
%%START
 -89\,\*\D(0)
 +76\,\*\D(1)
 +4\,\*\D(2)
 +12\,\*\Dd(-1,2)
 +42\,\*\Dd(0,2)
 +9\,\*\Dd(1,2) )
 \,-\, 72\,\*\z4\,\*\pgq(N)
\bigg)
%%STOP
\nn \\[1mm] & & \mbox{\hspn}
%%START
  + \frct{16}{27}\* \cf\* \ca\* \nfs \* \bigg(
 -4/3\,\*\th\,\*\St(-2)\,\*\D(-2)
 + \pgq(N) \* (
 -2\,\*\S(-4)
 -2\,\*\S(4)
 -23\,\*\Ss(-3,1)
%%STOP
\nn \\[-0.5mm] & & \mbox{\hspp}
%%START
 -20\,\*\Ss(-2,-2)
 -2\,\*\Ss(-2,2)
 -24\,\*\Ss(1,-3)
 +14\,\*\Ss(1,3)
 -12\,\*\Ss(2,-2)
 -8\,\*\Ss(2,2)
 +25\,\*\Ss(3,1)
 +20\,\*\Sss(-2,1,1)
%%STOP
\nn \\[0.5mm] & & \mbox{\hspp}
%%START
 +46\,\*\Sss(1,-2,1)
 +58\,\*\Sss(1,1,-2)
 +27\,\*\Sss(1,1,2)
 +25\,\*\Sss(1,2,1)
 +26\,\*\Sss(2,1,1)
 -16\,\*\Ssss(1,1,1,1) )
%%STOP
\nn \\[0.5mm] & & \mbox{\hspp}
%%START
 +\S(3)\,\*(-518/3\,\*\D(-1)
 +268/3\,\*\D(0)
 +106/3\,\*\D(1)
 +40/3\,\*\D(2)
 +32\,\*\Dd(-1,2)
 +112\,\*\Dd(0,2)
 +58\,\*\Dd(1,2))
%%STOP
\nn \\[0.5mm] & & \mbox{\hspp}
%%START
 +\Ss(2,1)\,\*(-461/3\,\*\D(-1)
 +427/3\,\*\D(0)
 -188/3\,\*\D(1)
 +4/3\,\*\D(2)
 -16\,\*\Dd(-1,2)
 +28\,\*\Dd(0,2)
 +27\,\*\Dd(1,2))
%%STOP
\nn \\[0.5mm] & & \mbox{\hspp}
%%START
 +\Ss(1,2)\,\*(-137\,\*\D(-1)
 +445/3\,\*\D(0)
 -248/3\,\*\D(1)
 -4/3\,\*\D(2)
 -24\,\*\Dd(-1,2)
 +12\,\*\Dd(0,2)
 +5\,\*\Dd(1,2))
%%STOP
\nn \\[0.5mm] & & \mbox{\hspp}
%%START
 +\Sss(1,1,1)\,\*(291\,\*\D(-1)
 -796/3\,\*\D(0)
 +299/3\,\*\D(1)
 -44/3\,\*\D(2)
 +52\,\*\Dd(-1,2)
 -140\,\*\Dd(0,2)
 -12\,\*\Dd(1,2))
%%STOP
\nn \\[0.5mm] & & \mbox{\hspp}
%%START
 +\Ss(1,-2)\,\*(-1306/3\,\*\D(-1)
 +808/3\,\*\D(0)
 +34/3\,\*\D(1)
 +16\,\*\D(2)
 +48\,\*\Dd(-1,2)
 +192\,\*\Dd(0,2)
 +130\,\*\Dd(1,2))
%%STOP
\nn \\[0.5mm] & & \mbox{\hspp}
%%START
 +\Ss(-2,1)\,\*(-914/3\,\*\D(-1)
 +416/3\,\*\D(0)
 +242/3\,\*\D(1)
 +16\,\*\D(2)
 +48\,\*\Dd(-1,2)
 +192\,\*\Dd(0,2)
 +118\,\*\Dd(1,2))
%%STOP
\nn \\[0.5mm] & & \mbox{\hspp}
%%START
 +\S(-3)\,\*(694/3\,\*\D(-1)
 -404/3\,\*\D(0)
 -50/3\,\*\D(1)
 -32/3\,\*\D(2)
 -32\,\*\Dd(-1,2)
 -112\,\*\Dd(0,2)
 -72\,\*\Dd(1,2))
%%STOP
\nn \\[0.5mm] & & \mbox{\hspp}
%%START
 +\S(2)\,\*(169/6\,\*\D(-1)
 -59/18\,\*\D(0)
 +277/18\,\*\D(1)
 +64/9\,\*\D(2)
 +284/3\,\*\Dd(-1,2)
 -127/3\,\*\Dd(0,2)
%%STOP
\nn \\[0.5mm] & & \mbox{\hspp}
%%START
 +24\,\*\Dd(0,3)
 -77/2\,\*\Dd(1,2)
 +\Dd(1,3))
 +\Ss(1,1)\,\*(-23147/36\,\*\D(-1)
 +19511/36\,\*\D(0)
 -1630/9\,\*\D(1)
%%STOP
\nn \\[0.5mm] & & \mbox{\hspp}
%%START
 +74/3\,\*\D(2)
 -128\,\*\Dd(-1,2)
 +709/2\,\*\Dd(0,2)
 -36\,\*\Dd(0,3)
 +1175/12\,\*\Dd(1,2)
 -28\,\*\Dd(1,3)
 -80/3\,\*\Dd(2,2))
%%STOP
\nn \\[0.5mm] & & \mbox{\hspp}
%%START
 +\S(-2)\,\*(5380/9\,\*\D(-1)
 -1520/3\,\*\D(0)
 +55/3\,\*\D(1)
 -376/9\,\*\D(2)
 -356/3\,\*\Dd(-1,2)
 -1148/3\,\*\Dd(0,2)
%%STOP
\nn \\[0.5mm] & & \mbox{\hspp}
%%START
 -80\,\*\Dd(0,3)
 -422/3\,\*\Dd(1,2)
 +138\,\*\Dd(1,3)
 +16/3\,\*\Dd(2,2))
 +\S(1)\,\*(7145/216\,\*\D(-1)
 +3775/36\,\*\D(0)
%%STOP
\nn \\[0.5mm] & & \mbox{\hspp}
%%START
 -3527/36\,\*\D(1)
 +6158/27\,\*\D(2)
 +1331/6\,\*\Dd(-1,2)
 +4099/36\,\*\Dd(0,2)
 +1523/6\,\*\Dd(0,3)
 +36\,\*\Dd(0,4)
%%STOP
\nn \\[0.5mm] & & \mbox{\hspp}
%%START
 -128\,\*\Dd(1,2)
 +1513/12\,\*\Dd(1,3)
 -87\,\*\Dd(1,4)
 +1532/9\,\*\Dd(2,2)
 -32/3\,\*\Dd(2,3))
 +304049/324\,\*\D(-1)
%%STOP
\nn \\[0.5mm] & & \mbox{\hspp}
%%START
 -66833/72\,\*\D(0)
 +28097/48\,\*\D(1)
 -57623/81\,\*\D(2)
 -19435/54\,\*\Dd(-1,2)
 -26555/36\,\*\Dd(0,2)
%%STOP
\nn \\[0.5mm] & & \mbox{\hspp}
%%START
 -8183/18\,\*\Dd(0,3)
 -356/3\,\*\Dd(0,4)
 -12\,\*\Dd(0,5)
 -17785/72\,\*\Dd(1,2)
 -6941/36\,\*\Dd(1,3)
 +836/3\,\*\Dd(1,4)
%%STOP
\nn \\[0.5mm] & & \mbox{\hspp}
%%START
 -88\,\*\Dd(1,5)
 -2632/9\,\*\Dd(2,2)
 +256/3\,\*\Dd(2,3)
 \,+\, \z3 \,\* ( \,
 2\,\*\Ndelta
 +36\,\*\pgq(N)\,\*\S(1)
 -206\,\*\D(-1)
%%STOP
\nn \\[0.5mm] & & \mbox{\hspp}
%%START
 +378\,\*\D(0)
 -324\,\*\D(1)
 -16\,\*\D(2)
 -48\,\*\Dd(-1,2)
 -168\,\*\Dd(0,2)
 -72\,\*\Dd(1,2))
 + \,72\,\*\z4\,\*\pgq(N)
\bigg)
%% ;
%%STOP
\label{eq:ggq3Nnf2}
\; .
\eea
The $\z4$ contributions to all 4-loop splitting functions in 
eq.~(\ref{eq:sgEvol}) have been derived in ref.~\cite{Davies:2017hyl}
from the no-$\pi^2$ theorem~\cite{Jamin:2017mul,Baikov:2018wgs}.
The very simple $\z4$ parts of eq.~(\ref{eq:ggq3Nnf2}) proportional
to 
\beq
\label{eq:pgqN}
  \pgq(N) \;=\;      2/(N-1) \,-\, 2/N  \,+\, 1/(N+1)
          \;\equiv\; 2\,\D(-1) - 2\,\D(0) \,+\, \D(1)
\; .
\eeq
agree with that result.
With the exception of the first term in the $C_F C_A\,\nfs$ part,
all harmonic sums are to be evaluated at~$N$; this argument has been 
suppressed in eq.~(\ref{eq:ggq3Nnf2}) for brevity. 

The presence of a contribution with $(N-2)^{-1}\, S_{-2}(N-2)$ 
-- which does, of course, not lead to a singularity at $N=2$ -- 
is clearly indicated by the presence of a $\delta(N\!-\!2) \,\z3$ term.
Such terms do not occur in the splitting functions to three loops,
but they are present in the coefficient functions for inclusive
DIS already at order $\as(2)$ \cite{vanNeerven:1991nn,Moch:1999eb}.
Very recently, the same structure was found to appear in the 
$C_F C_A\,\nfs$ part of the pure-singlet anomalous dimension 
$\gamma_{\,\rm ps}^{\,(3)}(N)$ \cite{Gehrmann:2023cqm}.

As expected from the values to $N \leq 60$, the overall weight 
(obtained for each term by adding the power of $1/(N+a)$ to the weight 
of the harmonic sum) reaches $w\!=\!6$ for the $\cfs\,\nfs$ part, 
and $w\!=\!5$ for the $C_F C_A\,\nfs$ contribution. 
Only non-alternating sums of $w\!=\!4$ enter the former, as expected 
from the $\cfs\,\nf$ 3-loop expression; and the alternating $w\!=\!3$ 
sums occur with a particular common prefactor in the second line of 
eq.~(\ref{eq:ggq3Nnf2}). 
The endpoint behaviour will be addressed below.

The inverse Mellin transform of eq.~(\ref{eq:ggq3Nnf2}) can be 
obtained by an algebraic procedure~\cite{Moch:1999eb,Remiddi:1999ew} 
based on the fact that harmonic sums occur as coefficients of the 
Taylor expansion of harmonic polylogarithms. 
This procedure results in
\bea
\lefteqn{ P_{gq}^{\,(3)}(x) \Big|_{\nfs} \;=\; }
  \nn \\[-1mm] & & \mbox{\hspn}
%%START
%%
%%L texPgq3cf2nf2 = 
  \frct{16}{27}\* \cfs\* \nfs \* \bigg( 
\ptmx\*\Big(
  156\,\*\H(5)
 +12\,\*\Hh(3,2)
 +60\,\*\Hh(4,0)
 +24\,\*\Hhh(3,0,0)
 +12\,\*\Hhh(3,1,0)
 +12\,\*\Hhh(3,1,1)
%%STOP
\nn \\[-1mm] & & \mbox{\hspp}
%%START
 -12\,\*\z2\,\*\H(3)
 -156\,\*\z2\,\*\Hhh(0,0,0)
 -36\,\*\z3\,\*\Hh(0,0)
 -24\,\*\z4\,\*\H(0)
 -12\,\*\z2\,\*\z3
 -12\,\*\z5\Big)
%%STOP
\nn \\[-1mm] & & \mbox{\hspp}
%%START
 +\pxgq\*\Big(
  8\,\*\Hh(1,3)
 +28\,\*\Hhh(1,1,2)
 +8\,\*\Hhh(1,2,0)
 +16\,\*\Hhh(1,2,1)
 -8\,\*\Hhhh(1,0,0,0)
 +16\,\*\Hhhh(1,1,0,0)
 +36\,\*\Hhhh(1,1,1,0)
%%STOP
\nn \\[-0.5mm] & & \mbox{\hspp}
%%START
 -16\,\*\Hhhh(1,1,1,1)
 -8\,\*\z2\,\*\Hh(1,0)
 -28\,\*\z2\,\*\Hh(1,1)
 +16\,\*\z3\,\*\H(1)
 \Big)
 +\H(4)\,\*(14
 -31\,\*x)
 -96\,\*\Hh(-3,0)
%%STOP
\nn \\[0.5mm] & & \mbox{\hspp}
%%START
 +\Hh(-2,2)\,\*(-288
 +96\,\*x^{-1}
 +144\,\*x)
 +\Hh(2,2)\,\*(-20
 -16\,\*x^{-1}
 +28\,\*x)
 +\Hh(3,0)\,\*(-80
 +130\,\*x)
%%STOP
\nn \\[0.5mm] & & \mbox{\hspp}
%%START
 +\Hh(3,1)\,\*(-102
 +81\,\*x)
 +\Hhh(-2,-1,0)\,\*(288
 -96\,\*x^{-1}
 -144\,\*x)
 +\Hhh(-2,0,0)\,\*(-192
 +64\,\*x^{-1}
%%STOP
\nn \\[0.5mm] & & \mbox{\hspp}
%%START
 +96\,\*x)
 +\Hhh(2,0,0)\,\*(-100
 -64\,\*x^{-1}
 -10\,\*x)
 +\Hhh(2,1,0)\,\*(-20
 -16\,\*x^{-1}
 +28\,\*x)
 +\Hhh(2,1,1)\,\*(-22
%%STOP
\nn \\[0.5mm] & & \mbox{\hspp}
%%START
 -16\,\*x^{-1}
 +29\,\*x)
 +\Hhhh(0,0,0,0)\,\*(-766
 +485\,\*x)
 +\H(3)\,\*(1335
 +390\,\*x
 +160/3\,\*x^2)
 +\Hh(-2,0)\,\*(480
%%STOP
\nn \\[0.5mm] & & \mbox{\hspp}
%%START
 -424/3\,\*x^{-1}
 -312\,\*x
 +64/3\,\*x^2)
 +\Hh(-1,2)\,\*(248\,\*x^{-1}
 -216\,\*x
 +32\,\*x^2)
 +\Hh(1,2)\,\*(418/3
%%STOP
\nn \\[0.5mm] & & \mbox{\hspp}
%%START
 -332/3\,\*x^{-1}
 -275/3\,\*x
 +16/3\,\*x^2)
 +\Hh(2,0)\,\*(1552/3
 +368/3\,\*x^{-1}
 -149/3\,\*x
 +112/3\,\*x^2)
%%STOP
\nn \\[0.5mm] & & \mbox{\hspp}
%%START
 +\Hh(2,1)\,\*(161/3
 +36\,\*x^{-1}
 -244/3\,\*x
 +32/3\,\*x^2)
 +\Hhh(-1,-1,0)\,\*(-248\,\*x^{-1}
 +216\,\*x
 -32\,\*x^2)
%%STOP
\nn \\[0.5mm] & & \mbox{\hspp}
%%START
 +\Hhh(-1,0,0)\,\*(496/3\,\*x^{-1}
 -144\,\*x
 +64/3\,\*x^2)
 +\Hhh(0,0,0)\,\*(563/3
 -91/3\,\*x
 +160/3\,\*x^2)
%%STOP
\nn \\[0.5mm] & & \mbox{\hspp}
%%START
 +\Hhh(1,0,0)\,\*(764/3
 -244\,\*x^{-1}
 -259/3\,\*x
 +64/3\,\*x^2)
 +\Hhh(1,1,0)\,\*(542/3
 -152\,\*x^{-1}
 -361/3\,\*x
%%STOP
\nn \\[0.5mm] & & \mbox{\hspp}
%%START
 +16/3\,\*x^2)
 +\Hhh(1,1,1)\,\*(-268/3
 +116\,\*x^{-1}
 +409/6\,\*x
 +16/3\,\*x^2)
 +\H(-2)\,\*(-144\,\*\z2\,\*x^{-1}
%%STOP
\nn \\[0.5mm] & & \mbox{\hspp}
%%START
 +432\,\*\z2
 -216\,\*\z2\,\*x)
 +\H(2)\,\*(-6727/9
 -3056/9\,\*x^{-1}
 -2891/18\,\*x
 -1384/9\,\*x^2
%%STOP
\nn \\[0.5mm] & & \mbox{\hspp}
%%START
 +64\,\*\z2\,\*x^{-1}
 +164\,\*\z2
 +44\,\*\z2\,\*x)
 +\Hh(-1,0)\,\*(-160
 -2408/9\,\*x^{-1}
 +56\,\*x
 -464/9\,\*x^2)
%%STOP
\nn \\[0.5mm] & & \mbox{\hspp}
%%START
 +\Hh(0,0)\,\*(-8110/3
 -4649/6\,\*x
 -2080/9\,\*x^2
 -14\,\*\z2
 +31\,\*\z2\,\*x)
 +\Hh(1,0)\,\*(-850/3
%%STOP
\nn \\[0.5mm] & & \mbox{\hspp}
%%START
 +3776/9\,\*x^{-1}
 +41/6\,\*x
 -896/9\,\*x^2)
 +\Hh(1,1)\,\*(-3461/18
 +491/9\,\*x^{-1}
 +2027/36\,\*x
%%STOP
\nn \\[0.5mm] & & \mbox{\hspp}
%%START
 -376/9\,\*x^2)
 +\H(-1)\,\*(-372\,\*\z2\,\*x^{-1}
 +324\,\*\z2\,\*x
 -48\,\*\z2\,\*x^2)
 +\H(0)\,\*(1344
%%STOP
\nn \\[0.5mm] & & \mbox{\hspp}
%%START
 +10528/27\,\*x^{-1}
 +389/36\,\*x
 +6494/27\,\*x^2
 -424/3\,\*\z2\,\*x^{-1}
 -1335\,\*\z2
 -702\,\*\z2\,\*x
%%STOP
\nn \\[0.5mm] & & \mbox{\hspp}
%%START
 -160/3\,\*\z2\,\*x^2
 +96\,\*\z3\,\*x^{-1}
 -48\,\*\z3
 +342\,\*\z3\,\*x)
 +\H(1)\,\*(72371/36
 -93103/54\,\*x^{-1}
%%STOP
\nn \\[0.5mm] & & \mbox{\hspp}
%%START
 -30551/72\,\*x
 +2678/27\,\*x^2
 +704/3\,\*\z2\,\*x^{-1}
 -418/3\,\*\z2
 -49/3\,\*\z2\,\*x
 -64/3\,\*\z2\,\*x^2)
%%STOP
\nn \\[0.5mm] & & \mbox{\hspp}
%%START
 -363779/216
 +368893/324\,\*x^{-1}
 +326947/432\,\*x
 -7090/81\,\*x^2
 +72\,\*\z2\,\*x^{-1}
%%STOP
\nn \\[0.5mm] & & \mbox{\hspp}
%%START
 +6727/9\,\*\z2
 +3899/18\,\*\z2\,\*x
 +1384/9\,\*\z2\,\*x^2
 -116/3\,\*\z3\,\*x^{-1}
 +674\,\*\z3
 -1288\,\*\z3\,\*x
%%STOP
\nn \\[0.5mm] & & \mbox{\hspp}
%%START
 +80/3\,\*\z3\,\*x^2
 +148\,\*\z4\,\*x^{-1}
 -1229/2\,\*\z4
 +1295/4\,\*\z4\,\*x
\bigg)
%%STOP
\nn \\[2mm] & & \mbox{\hspn}
%%START
+ \frct{16}{27}\* \cf\* \ca\* \nfs \* \bigg(
\pxgq\*\Big(
 -25\,\*\Hh(1,3)
 -20\,\*\Hhh(1,-2,0)
 +25\,\*\Hhh(1,1,2)
 +8\,\*\Hhh(1,2,0)
 +26\,\*\Hhh(1,2,1)
%%STOP
\nn \\[-1mm] & & \mbox{\hspp}
%%START
 -2\,\*\Hhhh(1,0,0,0)
 -14\,\*\Hhhh(1,1,0,0)
 +27\,\*\Hhhh(1,1,1,0)
 +16\,\*\Hhhh(1,1,1,1)
 +21\,\*\z2\,\*\Hh(1,0)
 +4\,\*\z2\,\*\Hh(1,1)
 -33\,\*\z3\,\*\H(1)
\Big)
%%STOP
\nn \\[-1mm] & & \mbox{\hspp}
%%START
 +\pmxgq\*\Big(
 -23\,\*\Hh(-1,3)
 +12\,\*\Hhh(-1,-2,0)
 +46\,\*\Hhh(-1,-1,2)
 -2\,\*\Hhh(-1,2,0)
 -20\,\*\Hhh(-1,2,1)
%%STOP
\nn \\[-1mm] & & \mbox{\hspp}
%%START
 -58\,\*\Hhhh(-1,-1,-1,0)
 +24\,\*\Hhhh(-1,-1,0,0)
 +2\,\*\Hhhh(-1,0,0,0)
 -75\,\*\z2\,\*\Hh(-1,-1)
 +19\,\*\z2\,\*\Hh(-1,0)
 +75\,\*\z3\,\*\H(-1)
 \Big)
%%STOP
\nn \\[-0.5mm] & & \mbox{\hspp}
%%START
 +\H(4)\,\*(36
 -16\,\*x)
 +80\,\*\Hh(-3,0)
 +\Hh(-2,2)\,\*(192
 -48\,\*x^{-1}
 -72\,\*x)
 +\Hh(2,2)\,\*(-28
 +16\,\*x^{-1}
 -2\,\*x)
%%STOP
\nn \\[0.5mm] & & \mbox{\hspp}
%%START
 +\Hh(3,0)\,\*(24
 +2\,\*x)
 +\Hh(3,1)\,\*(36
 +22\,\*x)
 +\Hhh(-2,-1,0)\,\*(-192
 +48\,\*x^{-1}
 +72\,\*x)
 +\Hhh(-2,0,0)\,\*(112
%%STOP
\nn \\[0.5mm] & & \mbox{\hspp}
%%START
 -32\,\*x^{-1}
 -48\,\*x)
 +\Hhh(2,0,0)\,\*(112
 +32\,\*x^{-1}
 +44\,\*x)
 +\Hhh(2,1,0)\,\*(-12
 +24\,\*x^{-1}
 +22\,\*x)
%%STOP
\nn \\[0.5mm] & & \mbox{\hspp}
%%START
 +\Hhh(2,1,1)\,\*(-140
 +52\,\*x^{-1}
 +4\,\*x)
 +\Hhhh(0,0,0,0)\,\*(12
 -6\,\*x)
 +\H(3)\,\*(-1499/6
 -343/2\,\*x
 +4/3\,\*x^2)
%%STOP
\nn \\[0.5mm] & & \mbox{\hspp}
%%START
 +\Hh(-2,0)\,\*(-956/3
 +212/3\,\*x^{-1}
 +152\,\*x
 -32/3\,\*x^2)
 +\Hh(-1,2)\,\*(-140/3
 -638/3\,\*x^{-1}
%%STOP
\nn \\[0.5mm] & & \mbox{\hspp}
%%START
 +242/3\,\*x
 -16\,\*x^2)
 +\Hh(1,2)\,\*(277/3
 -311/3\,\*x^{-1}
 -188/3\,\*x
 +4/3\,\*x^2)
 +\Hh(2,0)\,\*(67/3
%%STOP
\nn \\[0.5mm] & & \mbox{\hspp}
%%START
 -212/3\,\*x^{-1}
 -265/6\,\*x
 -4/3\,\*x^2)
 +\Hh(2,1)\,\*(525/2
 -76\,\*x^{-1}
 -7/4\,\*x
 -12\,\*x^2)
%%STOP
\nn \\[0.5mm] & & \mbox{\hspp}
%%START
 +\Hhh(-1,-1,0)\,\*(460/3
 +958/3\,\*x^{-1}
 -34/3\,\*x
 +16\,\*x^2)
 +\Hhh(-1,0,0)\,\*(-260/3
 -550/3\,\*x^{-1}
%%STOP
\nn \\[0.5mm] & & \mbox{\hspp}
%%START
 +50/3\,\*x
 -32/3\,\*x^2)
 +\Hhh(0,0,0)\,\*(-332/3
 -19/12\,\*x
 +8/3\,\*x^2)
 +\Hhh(1,0,0)\,\*(-184/3
%%STOP
\nn \\[0.5mm] & & \mbox{\hspp}
%%START
 +434/3\,\*x^{-1}
 -106/3\,\*x
 -40/3\,\*x^2)
 +\Hhh(1,1,0)\,\*(283/3
 -83\,\*x^{-1}
 -248/3\,\*x
 -4/3\,\*x^2)
%%STOP
\nn \\[0.5mm] & & \mbox{\hspp}
%%START
 +\Hhh(1,1,1)\,\*(700/3
 -259\,\*x^{-1}
 -299/3\,\*x
 +44/3\,\*x^2)
 +\H(-2)\,\*(72\,\*\z2\,\*x^{-1}
 -288\,\*\z2
 +108\,\*\z2\,\*x)
%%STOP
\nn \\[0.5mm] & & \mbox{\hspp}
%%START
 +\H(2)\,\*(16495/36
 +1067/6\,\*x^{-1}
 +610/9\,\*x
 +1202/9\,\*x^2
 -40\,\*\z2\,\*x^{-1}
 -68\,\*\z2
 -34\,\*\z2\,\*x)
%%STOP
\nn \\[0.5mm] & & \mbox{\hspp}
%%START
 +\Hh(-1,0)\,\*(226/3
 +4/3\,\*x^{-2}
 +1498/9\,\*x^{-1}
 +7/3\,\*x
 +232/9\,\*x^2)
 +\Hh(0,0)\,\*(8735/18
%%STOP
\nn \\[0.5mm] & & \mbox{\hspp}
%%START
 +3091/36\,\*x
 +440/9\,\*x^2
 -36\,\*\z2
 +16\,\*\z2\,\*x)
 +\Hh(1,0)\,\*(1363/18
 -305/6\,\*x^{-1}
 +301/18\,\*x
%%STOP
\nn \\[0.5mm] & & \mbox{\hspp}
%%START
 +52/9\,\*x^2)
 +\Hh(1,1)\,\*(-8747/36
 +12383/36\,\*x^{-1}
 +1498/9\,\*x
 -10\,\*x^2)
%%STOP
\nn \\[0.5mm] & & \mbox{\hspp}
%%START
 +\H(-1)\,\*(1117/3\,\*\z2\,\*x^{-1}
 +370/3\,\*\z2
 -259/3\,\*\z2\,\*x
 +24\,\*\z2\,\*x^2)
 +\H(0)\,\*(-10607/12
%%STOP
\nn \\[0.5mm] & & \mbox{\hspp}
%%START
 -4916/27\,\*x^{-1}
 -12139/72\,\*x
 -9944/27\,\*x^2
 +212/3\,\*\z2\,\*x^{-1}
 +1499/6\,\*\z2
 +647/2\,\*\z2\,\*x
%%STOP
\nn \\[0.5mm] & & \mbox{\hspp}
%%START
 -4/3\,\*\z2\,\*x^2
 -72\,\*\z3\,\*x^{-1}
 -72\,\*\z3
 -146\,\*\z3\,\*x)
 +\H(1)\,\*(-4495/6
 +132025/216\,\*x^{-1}
%%STOP
\nn \\[0.5mm] & & \mbox{\hspp}
%%START
 +3983/36\,\*x
 -6500/27\,\*x^2
 -56\,\*\z2\,\*x^{-1}
 -47/3\,\*\z2
 +205/3\,\*\z2\,\*x
 +20/3\,\*\z2\,\*x^2)
%%STOP
\nn \\[0.5mm] & & \mbox{\hspp}
%%START
 +30829/108
 -191575/648\,\*x^{-1}
 -91561/432\,\*x
 +27377/81\,\*x^2
 -205/18\,\*\z2\,\*x^{-1}
%%STOP
\nn \\[0.5mm] & & \mbox{\hspp}
%%START
 -16495/36\,\*\z2
 -589/9\,\*\z2\,\*x
 -1202/9\,\*\z2\,\*x^2
 +104\,\*\z3\,\*x^{-1}
 -752\,\*\z3
 +8587/12\,\*\z3\,\*x
%%STOP
\nn \\[0.5mm] & & \mbox{\hspp}
%%START
 +8\,\*\z3\,\*x^2
 -114\,\*\z4\,\*x^{-1}
 +331\,\*\z4
 -59/2\,\*\z4\,\*x
\bigg)
%% ;
%%STOP
\label{eq:Pgq3nf2}
\; , 
\eea
where we have suppressed the argument $x$ of the harmonic
polylogarithms (HPLs) $\,H_{m_1,...,m_w}(x)$ and used the 
leading-order function $\,\pxgq = 2\,x^{\,-1} - 2 + x\,$ 
to shorten the $w=4$ part of the expressions.
For chains of indices zero we employ the abbreviated notation 
\beq
\label{eq:habbr}
  H_{{\:\! \footnotesize\underbrace{0,\ldots ,0}_{\scriptstyle m} },\,
  \pm 1,\, {\footnotesize \underbrace{0,\ldots ,0}_{\scriptstyle n} },
  \, \pm 1,\, \ldots}(x) \; = \; H_{\pm (m+1),\,\pm (n+1),\, \ldots}(x)
  \:\: .
\eeq
Corresponding to the maximal weights in the $N$-space expression
(\ref{eq:ggq3Nnf2}), i.e., $w=6$ for the $\cfs\*\,\nfs$ part and 
$w=5$ for the $C_F C_A \*\,\nfs$ contribution, the respective $x$-space
results in eq.~(\ref{eq:Pgq3nf2}) include HPLs up to $w=5$ and $w=4$. 
The term corresponding to $\,(N\!-\!2)^{-1}\, S_{-2}(N\!-\!2)\,$
is $\,x^{\,-2}\, {\rm H}_{-1,0}(x)$.

Of particular interest are the logarithmically enhanced endpoint 
contributions in the high-energy (small-$x$) and threshold (large-$x$) 
limits.
In the former limit, the flavour-singlet splitting functions are 
dominated by the BFKL single-log enhancement of the $x^{\,-1}$ terms.
The subdominant $x^{\,0}$ contributions show a double-log enhancement. 
In both cases the leading $\nfs$ terms are of overall
next-to-next-to-leading logarithmic (NNLL) accuracy. 
Eq.~(\ref{eq:Pgq3nf2}) leads to
\bea
 P_{\rm gq}^{\,(3)}(x)\bigg|_{\nfs} &\!\!=\!\!&
%%START
%%
%%L texPgq3nf2xto0=
  \frac{1}{x}\,\*\Lz\,\,\* \bigg[ 
  \cfs\*\,\nfs \*
  \Big(\,\frac{168448}{729}
        -\frac{6784}{81}\,\*\z2
        +\frac{512}{9}\,\*\z3
  \Big)
%%STOP
\nn \\ & & \mbox{\hspp\hspp}
%%START
  \,+\, \cf\,\*\ca\,\*\nfs \* 
  \Big( - \frac{78080}{729}
        + \frac{3392}{81}\,\*\z2
        - \frac{128}{3}\,\*\z3
  \Big) 
  \bigg]
%%STOP
\nn \\[1mm] & & \mbox{\hspn}
%%START
 + \frac{1}{x} \,\* \bigg[ \,\cfs\,\*\nfs \*
   \Big(\,\frac{1475572}{2187}
         +\frac{128}{3}\,\*\z2
         -\frac{1856}{81}\,\*\z3
         +\frac{2368}{27}\,\*\z4\Big)
%%STOP
\nn \\[1mm] & & \mbox{\hspp\hspp}
%%START
  \,+\, \cf\,\*\ca\,\*\nfs \* 
  \Big( - \frac{384878}{2187}
        - \frac{1640}{243}\,\*\z2
        + \frac{1664}{27}\,\*\z3
        - \frac{608}{9}\,\*\z4
  \Big)
  \bigg]
%%STOP
\nn \\[1mm] & & \mbox{\hspn}
%%START
 +\Lzp^4\,\,\* 
   \bigg[
         - \frac{1532}{81}\,\* \cfs\,\*\nfs 
         + \frac{8}{27}\,\* \cf\,\*\ca\,\*\nfs
   \bigg]
%%STOP
\nn \\[1mm] & & \mbox{\hspn}
%%START
+\Lzp^3\,\,\* \bigg[ \,\cfs\,\*\nfs \*
   \Big(\,
          \frac{4120}{243}
        - \frac{832}{27}\,\*\z2
   \Big)
      \,- \frac{2848}{243}\* \cf\,\*\ca\,\*\nfs 
 \bigg]
%% ;
%%STOP
 \;+\; \ldots
\label{eq:Pgq3xto0}
\eea
where we have suppressed further terms for brevity.
The $\Lzp^4$ contribution, which arise from the $\Dd(0,5)$ terms in 
eq.~(\ref{eq:ggq3Nnf2}), agrees with the prediction in eq.~(5.11) of 
ref.~\cite{Davies:2022ofz}, the rest is new.

The off-diagonal splitting functions exhibit a double logarithmic 
enhancement also at large-$x$. In the present case it reads  
\bea
 P_{\rm gq}^{\,(3)}(x)\bigg|_{\nfs} &\!\!=\!\!&
%%START
%%
%%L texPgq3nf2xto1=
  \Lmzp^4\,\*\,
    \frac{32}{81}\, \* \cf\,\*\nfs \,\* (\ca - \cf )
  \,-\, 
  \Lmzp^3\,\*\bigg[\,
        \frac{2404}{243} \,\* \cfs\,\*\nfs 
      - \frac{2656}{243} \,\* \cf\,\*\ca\,\*\nfs 
      \bigg]
%%STOP      
\nn \\[0.5mm] & & \mbox{\hspn\hspn}
%%START
 +\Lmzp^2\,\* \bigg[ \,\cfs\,\*\nfs \*
  \Big( - \frac{8870}{243}
        - \frac{32}{3}\,\*\z2
  \Big)
  + \cf\,\*\ca\,\*\nfs \* 
  \Big(\, \frac{18536}{243}
        + \frac{16}{27}\,\*\z2
  \Big)
  \bigg]
%%STOP
\nn \\[1.5mm] & & \mbox{\hspn\hspn}
%%START
 +\Lmz\,\*\, \bigg[ \,\cfs\,\*\nfs \*
  \Big(\, \frac{1870}{81}
        - \frac{4144}{81}\,\*\z2
        - \frac{320}{27}\,\*\z3
  \Big)
%%STOP
\nn \\[0.5mm] & & \mbox{\hspp\hspp\hspp\hspp\hspp}
%%START
  + \cf\,\*\ca\,\*\nfs \* 
  \Big(\, \frac{12866}{81}
        - \frac{160}{81}\,\*\z2
        + \frac{880}{27}\,\*\z3
  \Big)
  \bigg]
%% ;
%%STOP
  \;+\; \ldots
\label{eq:Pgq3xto1}
\eea
The $\Lmzp^4$ contribution, including the terms suppressed by powers 
of $\xm1$ not shown here, agrees with the predictions in eqs.~(5.16) 
and (5.25) of ref.~\cite{Soar:2009yh}, 
see also refs.~\cite{Almasy:2010wn,Almasy:2015dyv}.
The remaining terms in eq.~(\ref{eq:Pgq3xto1}) were not known before.

To summarize, we have derived the $\nfs$ contributions to 
$P_{\,\rm gq}^{\,(3)}(x)$. Together with ref.~\cite{Gehrmann:2023cqm}
this completes the $\nfs$ parts of the 4-loop (N$^3$LO) 
quark-to-parton splitting functions for the evolution of unpolarized 
parton distributions of hadrons. 
As usual, our results refer to the \MSb\ scheme, written in terms of 
the expansion parameter $\ars \,=\,\als(\mus)/(4\pi)$, i.e., we have, 
without loss or information, identified the renormalization scale 
$\mu_{\:\!\sf r}$ with the factorization scale $\mu$ in 
eq.~(\ref{eq:Pgq3xto1}).

While the $\nfs$ parts are much larger, at any relevant number $\nf$ 
of light flavours, than the corresponding leading large-$\nf$ terms
-- see, e.g., the values at $N=8$ in eqs.~(17)$\,$-$\,$20 of ref.~%
\cite{Moch:2021qrk} --
these results are not yet of direct relevance for phenomenological
analyses of hard scattering processes.

It is not feasible to extend the present results to the complete 
N$^3$LO splitting functions with the method applied here. 
We expect it to be a formidable task also for other approaches 
based on the method of differential equations 
\cite{Blumlein:2017dxp,Ablinger:2018zwz,Blumlein:2022gpp,%
Gehrmann:2023ksf,Gehrmann:2023cqm,Basdew-Sharma:2022vya}.
Hence phenomenological analyses will have to rely on approximate
$x$-space expressions for the time being, as presented in 
ref.~\cite{Moch:2017uml} for the large-$\nc$ suppressed non-singlet
contributions and in 
refs.~\cite{Falcioni:2023luc,Falcioni:2023vqq,MRUVV-tba} for the
flavour-singlet cases. 
We hope to be able to present results based on the even moments 
$2 \leq N \leq 20$ also for the complete $P_{\,\rm gq}^{\,(3)}(x)$
and $P_{\,\rm gg}^{\,(3)}(x)$ in the near future.

The biggest issue in such $x$-space approximations, as also noted in 
ref.~\cite{Gehrmann:2023cqm}, are the hitherto unknown $x^{\,-1} 
\ln x$ contributions, i.e., the NNLL corrections in the high-energy
BFKL limit. 
With ref.~\cite{Gehrmann:2023cqm} and the present results, the first
two such contributions have been derived. These results should not 
only prove useful in the context of approximate expressions, but also
as checks of future calculations of NNLL corrections in the BFKL
limit and their transformation to the \MSb\ scheme, which in itself
will be a non-trivial operation, see 
refs.~\cite{Ciafaloni:2005cg,Ciafaloni:2006yk}.
Also the results for the double-logarithmically enhanced contributions
in the small-$x$ and large-$x$ limits can provide input for the
extension of the respective resummations to a higher logarithmic 
accuracy.

\vspace{2mm}
{\sc Form} files with our results have been deposited at the preprint 
server {\tt https://arXiv.org} with the sources of this letter.
They are also available from the authors upon request.

% ---------------------------------------------------------------------
%
\subsection*{Acknowledgements}
\vspace*{-1mm}

G.F. would like to acknowledge the influence of Marcello Ciafaloni, who
passed away recently, on his decision to work on quantum field theory.
This work has been supported by
the Vidi grant 680-47-551 of the Dutch Research Council (NWO),
the UKRI FLF Mr/S03479x/1;
the Consolidated Grant ST/P0000630/1 {\it Particle Physics at the Higgs
Centre} of the UK Science and Technology Facilities Council (STFC);
the ERC Starting Grant 715049 {\it QCDforfuture},
the EU's Horizon Europe research and innovation programme under the 
Marie Sklodowska-Curie grant 101104792, {\it QCDchallenge};
the Deutsche Forschungsgemeinschaft through the Research Unit FOR 2926,
{\it Next Generation pQCD for Hadron Structure: Preparing for the EIC},
project number 40824754, and DFG grant MO~1801/4-2,
the ERC Advanced Grant 101095857 {\it Conformal-EIC};
and the STFC Consolidated Grant ST/T000988/1.
G.F. and S.M. are grateful to the Galileo Galilei Institute in Florence 
for hospitality and support during the scientific program on {\it 
Theory Challenges in the Precision Era of the Large Hadron Collider}, 
where part of this work was done.

%%{\footnotesize
{\small
\addtolength{\baselineskip}{-1.5mm}

\providecommand{\href}[2]{#2}\begingroup\raggedright\endgroup

}

\begin{thebibliography}{10}

\bibitem{Davies:2016jie}
J.~Davies, A.~Vogt, B.~Ruijl, T.~Ueda and J.A.M. Vermaseren,
% \emph{Large-$\nf$ contributions to the four-loop splitting functions in
% QCD}, 
 \href{https://doi.org/10.1016/j.nuclphysb.2016.12.012}{Nucl.\ Phys.\ B915
 (2017) 335},
\\[1mm]
 \href{https://arxiv.org/abs/1610.07477}{{\tt arXiv:1610.07477}}

\bibitem{Gehrmann:2023cqm}
T.~Gehrmann, A.~von Manteuffel, V.~Sotnikov and T.Z.~Yang,
% \emph{Complete $\nfs$ contributions to four-loop pure-singlet splitting 
% functions},
  \href{https://arxiv.org/abs/2308.07958}{{\tt arXiv:2308.07958}}

\bibitem{Moch:2017uml}
S.~Moch, B.~Ruijl, T.~Ueda, J. Vermaseren and A.~Vogt,
% \emph{{Four-Loop Non-Singlet Splitting Functions in the Planar Limit 
% and Beyond}},
 \href{https://doi.org/10.1007/JHEP10(2017)041}{JHEP 10 (2017) 041},
 \href{https://arxiv.org/abs/1707.08315}{{\tt arXiv:1707.08315}}

\bibitem{Falcioni:2022fdm}
G.~Falcioni and F.~Herzog,
% \emph{{Renormalization of gluonic leading-twist operators in covariant 
% gauges}},
 \href{https://doi.org/10.1007/JHEP05(2022)177}{JHEP 05 (2022) 177},
 \href{https://arxiv.org/abs/2203.11181}{{\tt arXiv:2203.11181}}

\bibitem{Gehrmann:2023ksf}
T.~Gehrmann, A.~von Manteuffel and T.-Z. Yang,
% \emph{{Renormalization of twist-two operators in covariant gauge to 
% three loops in QCD}},
 \href{https://doi.org/10.1007/JHEP04(2023)041}{JHEP 04 (2023) 041},
 \href{https://arxiv.org/abs/2302.00022}{{\tt arXiv:2302.00022}}

\bibitem{Falcioni:2023luc}
G.~Falcioni, F.~Herzog, S.~Moch and A.~Vogt,
% \emph{{Four-loop splitting functions in QCD - The quark-quark case -}},
 \href{https://doi.org/10.1016/j.physletb.2023.137944}{Phys.\ Lett.\ B842
 (2023) 137944},
 \href{https://arxiv.org/abs/2302.07593}{{\tt arXiv:2302.07593}}

\bibitem{Falcioni:2023vqq}
G.~Falcioni, F.~Herzog, S.~Moch and A.~Vogt,
% \emph{{Four-loop splitting functions in QCD -- The gluon-to-quark case}},
 Phys.\ Lett.\ B, to appear,
 \href{https://arxiv.org/abs/2307.04158}{{\tt arXiv:2307.04158}}

\bibitem{Moch:2021qrk}
S.~Moch, B.~Ruijl, T.~Ueda, J.A.M. Vermaseren and A.~Vogt,
% \emph{{Low moments of the four-loop splitting functions in QCD}},
  \href{https://doi.org/10.1016/j.physletb.2021.136853}{Phys. Lett. B825
  (2022) 136853},
\\[1mm]
  \href{https://arxiv.org/abs/2111.15561}{{\tt arXiv:2111.15561}}

\bibitem{MRUVV-tba}
S.~Moch, B.~Ruijl, T.~Ueda, J.A.M. Vermaseren and A.~Vogt, 
  \emph{{ to appear}}

\bibitem{Herzog:2016qas}
F.~Herzog, B.~Ruijl, T.~Ueda, J. Vermaseren, A.~Vogt,
% \emph{{FORM, Diagrams and Topologies}},
 \href{https://doi.org/10.22323/1.260.0073}{PoS LL2016 (2016) 073},
 \href{https://arxiv.org/abs/1608.01834}{{\tt arXiv:1608.01834}}

\bibitem{Vermaseren:2000nd}
J.~A.~M. Vermaseren,
  \emph{{New features of FORM}},
 \href{https://arxiv.org/abs/math-ph/0010025}{{\tt math-ph/0010025}}

\bibitem{Kuipers:2012rf}
J.~Kuipers, T.~Ueda, J. Vermaseren and J.~Vollinga,
% \emph{{FORM version 4.0}}, 
 \href{https://doi.org/10.1016/j.cpc.2012.12.028}{Comput. Phys. Comm.
 184 (2013) 1453},
\\[1mm]
 \href{https://arxiv.org/abs/1203.6543}{{\tt arXiv:1203.6543}}

\bibitem{Ruijl:2017dtg}
B.~Ruijl, T.~Ueda and J.~Vermaseren,
  \emph{{FORM version 4.2}},
 \href{https://arxiv.org/abs/1707.06453}{{\tt arXiv:1707.06453}}

\bibitem{vanRitbergen:1998pn}
T.~van Ritbergen, A.N. Schellekens and J.A.M. Vermaseren,
% \emph{{Group theory factors for Feynman diagrams}},
 \href{https://doi.org/10.1142/S0217751X99000038}{Int. J. Mod. Phys.
 A14 (1999) 41},
\\[1mm]
 \href{https://arxiv.org/abs/hep-ph/9802376}{{\tt hep-ph/9802376}}

\bibitem{Ruijl:2017cxj}
B.~Ruijl, T.~Ueda and J. Vermaseren,
% \emph{{Forcer, a FORM program for the parametric reduction of four-loop 
% massless propagator diagrams}},
 \href{https://doi.org/10.1016/j.cpc.2020.107198}
 {Comput. Phys. Comm. 253 (2020) 107198},
 \href{https://arxiv.org/abs/1704.06650}{{\tt arXiv:1704.06650}}

\bibitem{Gorishnii:1989gt}
S.G.~Gorishnii, S.A.~Larin, L.R.~Surguladze, F.V.~Tkachov,
% {\it Mincer: Program for Multiloop Calculations in Quantum Field Theory
% for the Schoonschip System},
  \href{https://doi.org/10.1016/0010-4655(89)90134-3}
  {Comput.\ Phys.\ Commun.\ 55 (1989) 381}

\bibitem{Larin:1991fz}
S.A.~Larin, F.V.~Tkachov and J.A.M.~Vermaseren,
  \href{https://www.nikhef.nl/~t68/FORMapplications/Mincer}
  {{\it The FORM version of MINCER}}, NIKHEF-H-91-18

\bibitem{Moch:2014sna}
S.~Moch, J.A.M.~Vermaseren and A.~Vogt,
% \emph{{The Three-Loop Splitting Functions in QCD: The Helicity-Dependent 
% Case}},
  \href{https://doi:10.1016/j.nuclphysb.2014.10.016}
  {Nucl. Phys. B889 (2014) 351}
  \href{https://arxiv.org/abs/1409.5131}{{\tt arXiv:1409.5131}}

\bibitem{Vermaseren:1998uu}
J.A.M. Vermaseren,
% \emph{{Harmonic sums, Mellin transforms and integrals}},
 \href{https://doi.org/10.1142/S0217751X99001032}{Int. J. Mod. Phys. A14
 (1999) 2037},
 \href{https://arxiv.org/abs/hep-ph/9806280}{{\tt hep-ph/9806280}}

\bibitem{Vogt:2004mw}
A.~Vogt, S.~Moch and J.A.M. Vermaseren,
% \emph{{The three-loop splitting functions in QCD: The singlet case}},
 \href{https://doi.org/10.1016/j.nuclphysb.2004.04.024}{Nucl. Phys. B691
 (2004) 129},
 \href{https://arxiv.org/abs/hep-ph/0404111}{{\tt hep-ph/0404111}}

\bibitem{axbAlg}
K.~Matthews,
  \href{http://www.numbertheory.org/PDFS/ax=b.pdf}{ 
  \emph{Solving $ax = b$ using the hermite normal form} (unpublished)}
  
\bibitem{Calc}
  \href{http://www.numbertheory.org/calc/krm_calc.html}
  {{\tt http://www.numbertheory.org/calc/krm\_calc.html}}

\bibitem{Lenstra1982}
A.K.~Lenstra, H.W.~Lenstra and L.~Lov{\'a}sz, 
% \emph{Factoring polynomials with rational coefficients},
  \href{https://dx.doi.org/10.1007/BF01457454}{Mathematische Annalen 
  261 (1982) 515}

\bibitem{Velizhanin:2012nm}
V.N.~Velizhanin, 
% \emph{{Three loop anomalous dimension of the non-singlet
% transversity operator in QCD}},
  \href{https://dx.doi.org/10.1016/j.nuclphysb.2012.06.010}
  {Nucl.~Phys.  B864 (2012) 113},
  \href{https://arxiv.org/abs/1203.1022}{{\tt arXiv:1203.1022}}

\bibitem{Davies:2017hyl}
J.~Davies and A.~Vogt,
% \emph{{Absence of $\pi^2$ terms in physical anomalous dimensions in DIS: 
% Verification and resulting predictions}},
 \href{https://doi.org/10.1016/j.physletb.2017.11.036}{Phys. Lett. B776
 (2018) 189--194},
 \href{https://arxiv.org/abs/1711.05267}{{\tt arXiv:1711.05267}}

\bibitem{Jamin:2017mul}
M.~Jamin and R.~Miravitllas, 
% \emph{{Absence of even-integer $\zeta$-function
% values in Euclidean physical quantities in QCD}},
  \href{http://dx.doi.org/10.1016/j.physletb.2018.02.030}
  {Phys. Lett. B779 (2018) 452}, 
  \href{http://arxiv.org/abs/1711.00787}{{\tt arXiv:1711.00787}}

\bibitem{Baikov:2018wgs}
P.~A. Baikov and K.~G. Chetyrkin, 
% \emph{{The structure of generic anomalous
% dimensions and no-$\pi$ theorem for massless propagators}},
  \href{http://dx.doi.org/10.1007/JHEP06(2018)141}{JHEP 06 (2018) 141}, 
  \href{http://arxiv.org/abs/1804.10088}{{\tt arXiv:1804.10088}}

\bibitem{vanNeerven:1991nn}
W.L.~van Neerven and E.B.~Zijlstra,
% \emph{{Order $\as(2)$ contributions to the deep inelastic Wilson 
% coefficient}},
  \href{https://dx.doi.org/10.1016/0370-2693(91)91024-P}
  {Phys. Lett. B272 (1991) 127}

\bibitem{Moch:1999eb}
S.~Moch and J.A.M.~Vermaseren, 
% \emph{{Deep inelastic structure functions at two loops}}, 
  \href{https://dx.doi.org/10.1016/S0550-3213(00)00045-6}{Nucl. Phys. B573 
  (2000) 853},
  \href{https://arxiv.org/abs/hep-ph/9912355}{{\tt hep-ph/9912355}}

\bibitem{Remiddi:1999ew}
E.~Remiddi and J.A.M.~Vermaseren,
% \emph{{Harmonic polylogarithms}},
  \href{https://dx.doi.org/10.1142/S0217751X00000367}
    {Int.\ J.\ Mod.\ Phys.\ A15 (2000) 725}, 
  \href{https://arxiv.org/abs/hep-ph/9905237}{{\tt hep-ph/9905237}}

\bibitem{Davies:2022ofz}
J.~Davies, C.H. Kom, S.~Moch and A.~Vogt,
% \emph{{Resummation of small-x double logarithms in QCD: inclusive 
% deep-inelastic scattering}},
 \href{https://doi.org/10.1007/JHEP08(2022)135}{JHEP 08 (2022) 135},
 \href{https://arxiv.org/abs/2202.10362}{{\tt arXiv:2202.10362}}

\bibitem{Soar:2009yh}
G.~Soar, S.~Moch, J. Vermaseren and A.~Vogt,
% \emph{{On Higgs-exchange DIS, physical evolution kernels and fourth-order 
% splitting functions at large x}}, 
 \href{https://doi.org/10.1016/j.nuclphysb.2010.02.003}{Nucl. Phys. B832
 (2010) 152},
 \href{https://arxiv.org/abs/0912.0369}{{\tt arXiv:0912.0369}}

\bibitem{Almasy:2010wn}
A.A.~Almasy, G.~Soar and A.~Vogt,
% \emph{{Generalized double-logarithmic large-x resummation in inclusive 
% deep-inelastic scattering}},
 \href{https://doi.org/10.1007/JHEP03(2011)030}{JHEP 03 (2011) 030},
 \href{https://arxiv.org/abs/1012.3352}{{\tt arXiv:1012.3352}}

\bibitem{Almasy:2015dyv}
A.A.~Almasy, N.A.~Lo Presti and A.~Vogt,
% \emph{{Generalized threshold resummation in inclusive DIS and 
% semi-inclusive electron-positron annihilation}},
  \href{https://doi.org/10.1007/JHEP01(2016)028}{JHEP 01 (2016) 028},
  \href{https://arxiv.org/abs/1511.08612}{{\tt arXiv:1511.08612}}

\bibitem{Blumlein:2017dxp}
J.~Bl\"umlein and C.~Schneider,
%  \emph{{The Method of Arbitrarily Large Moments to Calculate Single Scale 
%  Processes in Quantum Field Theory}},
   \href{https://doi.org/10.1016/j.physletb.2017.05.001}
   {Phys. Lett. B771 (2017) 31},
   \href{https://arxiv.org/abs/1701.04614}{{\tt arXiv:1701.04614}}

\bibitem{Ablinger:2018zwz}
J.~Ablinger, J.~Bl\"umlein, P.~Marquard, N.~Rana and C.~Schneider,
%  \emph{{Automated Solution of First Order Factorizable Systems of 
%  Differential Equations in One Variable}},
   \href{https://doi.org/10.1016/j.nuclphysb.2018.12.010} 
   {Nucl. Phys. B939 (2019) 253},
\\[1mm]
   \href{https://arxiv.org/abs/1810.12261}{{\tt arXiv:1810.12261}}

\bibitem{Blumlein:2022gpp}
J.~Bl\"umlein, P.~Marquard, C.~Schneider and K.~Sch\"onwald,
%  emph{{The massless three-loop Wilson coefficients for the 
%  deep-inelastic structure functions $F_2$, $F_L$, $xF_3$ and $g_1$}},
   \href{https://doi.org/10.1007/JHEP11(2022)156}
   {JHEP 11 (2022) 156},
   \href{https://arxiv.org/abs/2208.14325}{{\tt arXiv:2208.14325}}

\bibitem{Basdew-Sharma:2022vya}
A.~Basdew-Sharma, A.~Pelloni, F.~Herzog and A.~Vogt,
%  \emph{{Four-loop large-$\nf$ contributions to the non-singlet 
%  structure functions $F_{2}$ and $F_{L}$}},
   \href{https://doi.org/10.1007/JHEP03(2023)183}{JHEP 03 (2023) 183},
   \href{https://arxiv.org/abs/2211.16485}{{\tt arXiv:2211.16485}}

\bibitem{Ciafaloni:2005cg}
M.~Ciafaloni and D.~Colferai,
%  \emph{{Dimensional regularisation and factorisation schemes in the 
%  BFKL equation at subleading level}},
   \href{https://doi.org/10.1088/1126-6708/2005/09/069}{JHEP 09 (2005) 069},
   \href{https://arxiv.org/abs/hep-ph/0507106}{{\tt hep-ph/0507106}}

\bibitem{Ciafaloni:2006yk}
M.~Ciafaloni, D.~Colferai, G.P.~Salam, A.M.~Stasto,
%  \emph{{Minimal subtraction vs.~physical factorisation schemes in 
%  small-x QCD}},
   \href{https://doi.org/10.1016/j.physletb.2006.03.014}
   {Phys. Lett. B635 (2006) 320},
   \href{https://arxiv.org/abs/hep-ph/0601200}{{\tt hep-ph/0601200}}  

\end{thebibliography}
\end{document}